\documentclass[aps,prl,twocolumn,showpacs]{revtex4}

\usepackage{graphicx}% Include figure files
\usepackage{amsfonts}
\input epsf
\input{epsf.sty}
\begin{document}

\preprint{APS/123-QED}

\title{%
Competition between unconventional superconductivity and incommensurate antiferromagnetic order in CeRh$_{1-x}$Co$_{x}$In$_5$
}%

\author{S.~Ohira-Kawamura$^1$}
\email{kawamura.seiko@ocha.ac.jp}
\author{H.~Shishido$^2$}
\email{shishido@scphys.kyoto-u.ac.jp}
\author{A.~Yoshida$^3$, R.~Okazaki$^2$, H.~Kawano-Furukawa$^3$, 
  T.~Shibauchi$^2$, H.~Harima$^4$}
\author{Y.~Matsuda$^{2,5}$}

\affiliation{%
  $^1$Academic and Information Board, Ochanomizu University,
  Bunkyo-ku, Tokyo 112-8610, Japan\\
  $^2$Department of Physics, Kyoto University,
  Sakyo-ku, Kyoto 606-8502, Japan\\
  $^3$Department of Physics, Ochanomizu University,
  Bunkyo-ku, Tokyo 112-8610, Japan\\
  $^4$Department of Physics, Kobe University,
  Kobe 657-8501, Japan\\
  $^5$Institute for Solid State Physics, University of Tokyo,
  Kashiwa, Chiba 277-8581, Japan
}%

\date{\today}

\begin{abstract}

Elastic neutron diffraction measurements were performed on the
quasi-two dimensional heavy fermion system CeRh$_{1-x}$Co$_x$In$_5$,
ranging from an incommensurate antiferromagnet for low $x$ 
to an unconventional superconductor on the Co-rich end of the phase
diagram. We found that the superconductivity competes with the
incommensurate antiferromagnetic (AFM) order characterized by
{\boldmath $q$}$_I=(\frac{1}{2}, \frac{1}{2}, \delta)$ with
$\delta=0.298$, while it coexists with the commensurate AFM order with
{\boldmath $q$}$_c=(\frac{1}{2}, \frac{1}{2},\frac{1}{2})$\@. This is
in sharp contrast to the CeRh$_{1-x}$Ir$_x$In$_5$ system, where both
the commensurate and incommensurate magnetic orders coexist with the
superconductivity. These results reveal that particular areas on the Fermi surface nested by {\boldmath $q$}$_I$ play an active role in forming the superconducting state in CeCoIn$_5$.

\end{abstract}

\pacs{74.70.Tx, 74.25.Dw, 75.25.+z}%
%------------------------------------------------------------------%
% 74.25.Dw  Superconductivity phase diagrams
% 74.70.Tx  Heavy-fermion superconductors
% 75.25.+z  Spin arrangements in magnetically ordered materials
%		(including neutron and spin-polarized electron
%		studies, synchrotron-source X-ray scattering, etc.)
\maketitle

%------------Introduction------------%

The unconventional superconducting (SC) pairing state realized in strongly correlated electron systems, including heavy fermion and high-$T_c$ cuprates, often develops, to varying extents, in the proximity of a magnetically ordered state. Therefore, it is widely believed that the magnetic fluctuations play important roles for the Cooper pairing.  In fact, strong coupling between the magnetic excitation spectra and SC order parameter has been reported in high-$T_c$ cuprates~\cite{fon95} and a heavy fermion compound~\cite{sat01}\@.  To obtain further insights into the microscopic mechanism of unconventional superconductivity,  more detailed information of the electronic structure, especially the position on the Fermi surface which plays an active role for the pairing formation,  is strongly required.   In the case of high-$T_c$ cuprates with a simple two-dimensional (2D) Fermi surface,  the hot spots, at which the scattering rate is dramatically enhanced,  appear at certain parts of the Fermi surface, as a consequence of the strong 2D antiferromagnetic (AFM) fluctuations due to the nesting of the Fermi surface~\cite{dam03}\@.   On the other hand,  such information is still lacking in heavy fermion systems to date, mainly because the complicated 3D Fermi surface often makes it difficult 
to specify
%to know 
the actual active position on the Fermi surface.

A new heavy fermion family of Ce$M$In$_5$, where $M$ can be either Ir, Co or Rh,  has attracted much interest on account of the relationship between the superconductivity and magnetism~\cite{heg00,pet01}\@. 
CeCoIn$_5$ and CeIrIn$_5$ are superconductors with the SC transition temperatures of $T_c =2.3$~K and 0.4~K, respectively. 
The presence of strong AFM fluctuations associated with the quantum critical point (QCP) nearby has been reported in the normal state of CeCoIn$_5$~\cite{koh01,yas04}\@.  This, together with the $d$-wave (presumably $d_{x^2-y^2}$) gap symmetry~\cite{iza01},  indicates importance of the AFM fluctuations for the superconductivity of CeCoIn$_5$\@. 
On the other hand, in another compound CeRhIn$_5$, the superconductivity is highly suppressed, and an AFM order appears below $T_{\rm N}=3.8$~K\@. The propagating vector is determined to be {\boldmath $q$}$_I=(\frac{1}{2},\frac{1}{2},\delta)$, $\delta =0.297$, that is incommensurate with a tetragonal crystal lattice, by neutron diffraction measurements~\cite{bao00}\@. 
Furthermore, a new commensurate AFM order with {\boldmath $q$}$_c=(\frac{1}{2},\frac{1}{2},\frac{1}{2})$ has been found in  CeRh$_{1-x}$Ir$_x$In$_5$~\cite{chr05}, and then the superconductivity coexists with the two distinct magnetic orders in  a wide composition range ($0.25 \le x \le 0.6$)\@. 
Very recent neutron diffraction measurements further reported a similar coexistence even in CeRh$_{0.6}$Co$_{0.4}$In$_5$~\cite{yok06}\@.  Such an unusual coexistence of three different types of cooperative ordered states is quite unique among the unconventional superconductors. 
Then, it is important to understand their magnetic properties
for elucidating the mechanism of the unconventional superconductivity in the Ce$M$In$_5$ systems.

We here report the results of the elastic neutron diffraction measurements on CeRh$_{1-x}$Co$_x$In$_5$, ranging from the AFM metallic to the unconventional SC states.  We found that, in sharp contrast to CeRh$_{1-x}$Ir$_x$In$_5$,  the superconductivity is strongly suppressed by the incommensurate AFM order characterized by {\boldmath $q$}$_I=(\frac{1}{2},\frac{1}{2},0.298)$, while it coexists with the commensurate AFM order with {\boldmath $q$}$_c=(\frac{1}{2},\frac{1}{2},\frac{1}{2})$\@. 
These results provide important information of the positions on the Fermi surface which are responsible for the unconventional superconductivity in CeRh$_{1-x}$Co$_x$In$_5$\@. This is the first report to provide such information in the heavy fermion compounds. 
We will also discuss a difference in the SC states of CeCoIn$_5$ and CeIrIn$_5$, based on the different types of the coexistence of the magnetism and superconductivity.

%------------Experimental------------%

Single crystals of CeRh$_{1-x}$Co$_x$In$_5$ for $x=0$, 0.2, 0.3, 0.4, 0.6, 0.7, 0.75 and 1 were prepared by the self-flux method~\cite{shi02}\@. 
%The plate-shape samples with the sizes of  for $x \le 0.4$ and those of $\sim 3 \times 3 \times 0.5$~mm$^3$ for $x \ge 0.6$ were used for the neutron experiments. 
Elastic neutron diffraction experiments were carried out on the $x=0.3$, 0.4, 0.6, 0.7 and 0.75 samples at the triple-axis spectrometer GPTAS (4G) installed at the JRR-3 reactor in Japan Atomic Energy Agency\@. 
The samples with the typical size of $\sim 5 \times 5 \times 0.5$~mm$^3$ were
set with ($h~h~l$) scattering plane, and were cooled down to 0.7~K\@.
% by using a $^3$He cryostat. 
%The $x=0.3$, 0.4 and 0.6 samples were cooled down to 1.5~K by using a pumped $^4$He cryostat,and the $x=0.7$ and 0.75 ones down to 0.7~K with a 1K cryostat. 
%The scattering plane ($hhl$) was selected.  
The neutrons with momentum of $k=3.814$~\AA$^{-1}$ or 2.67~\AA$^{-1}$ were used for the measurements. The 40'-40'-40'-80' collimators and two pyrolytic graphite filters, which eliminate the higher-order reflections, were used. 
To check the sample quality, we have also measured the specific heat
%, magnetic susceptibility 
and resistivity of the samples 
%%%
with the same compositions as those used for the neutron diffraction measurements (the same batch) and the $x=0$, 0.2 and 1 samples.
%%%
%in the same batch.

%------------Results and Discussion------------%

% Fig. 1
\begin{figure}
  \includegraphics[scale=0.52]{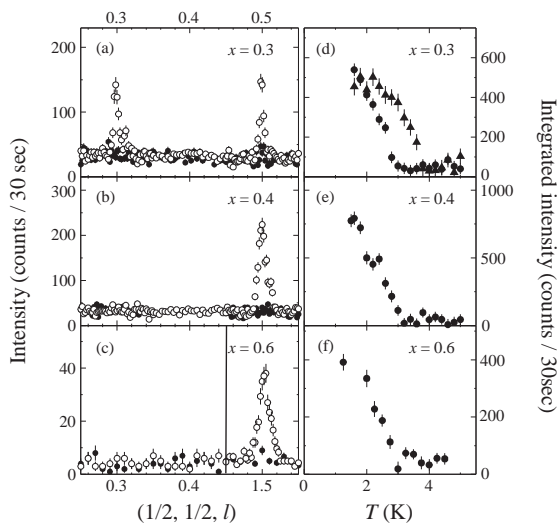}
  \caption{\label{neutron}
    Neutron diffraction profiles for CeRh$_{1-x}$Co$_x$In$_5$ for (a)~$x=0.3$,
    (b)~0.4 and (c)~0.6 for {\boldmath $Q$}$=(\frac12, \frac12, l)$\@.
    For $x=0.6$, the profile in the range 
    $1.45 \le l \le 1.55$ is shown instead of $0.45 \le l \le 0.55$, 
    because mosaic peaks are observed at the latter position.  
    Open and closed circles respectively show the results at 1.5~K and 5~K\@.
    Temperature dependences of the integrated intensities for the Bragg
    peaks at {\boldmath $q$}$_c =(\frac12, \frac12, \frac12)$
    (circles) and {\boldmath $q$}$_I =(\frac12, \frac12, 0.298)$ (triangles), 
    are obtained from longitudinal scans for (d)~$x=0.3$, (e)~0.4 and (f)~0.6.
  }
\end{figure}

Figures~\ref{neutron}(a), (b) and (c) respectively show the neutron diffraction profiles along the $l$ direction ({\boldmath $Q$}$= (\frac12, \frac12, l)$) for CeRh$_{1-x}$Co$_x$In$_5$ with $x=0.3$, 0.4 and 0.6 at $T=1.5$~K ($T < T_{\rm N}$) and 5~K ($T > T_{\rm N}$)\@. 
Magnetic Bragg peaks are observed at 1.5~K at {\boldmath $q$}$_c = (\frac{1}{2}, \frac{1}{2}, \frac{1}{2})$ for $x=0.3$, 0.4 and 0.6, indicating appearance of commensurate AFM orders, while such a peak is not observed for $x \ge 0.7$ (not shown) at least down to 0.7~K\@.  For $x=0.3$, in addition to the commensurate AFM peak, another magnetic Bragg peak is observed at {\boldmath $q$}$_I=(\frac12, \frac12, 0.298)$, which indicates an incommensurate AFM order.  Note that a similar incommensurate AFM peak is observed in CeRhIn$_5$.

Temperature dependences of the integrated intensities of the Bragg peaks for $x=0.3$, 0.4 and 0.6 are depicted in Figs.~\ref{neutron}(d), (e) and (f), respectively.  The commensurate AFM Bragg peaks (filled circles) develop below 3.0~K, 3.1~K, and 2.8~K for $x=0.3$, 0.4, and 0.6.  The incommensurate AFM Bragg peak with {\boldmath $q$}$_I = (\frac{1}{2}, \frac{1}{2},  0.298)$ appears below 3.7~K for $x=0.3$ (filled triangles)\@. 
%%% 2007-4-2
The averaged magnetic moments $M_c$ for the commensurate AFM order and $M_I$ for the incommensurate one are evaluated to be $M_c=0.28(2)$~$\mu_{\rm B}$/Ce and $M_I \sim 0.4$~$\mu_{\rm B}$/Ce for $x=0.3$, $M_c=0.31(4)$~$\mu_{\rm B}$/Ce for $x=0.4$ and $M_c=0.30(2)$~$\mu_{\rm B}$/Ce for $x=0.6$\@.
%%% SK added a ref.
For evaluating the moments, we assumed the spins lying on the basal plane, similar to the helical AFM moments in CeRhIn$_5$~\cite{bao00}\@. 
%The incommensurate moment for $x$=0.3 is $M_I \simeq 0.8~\mu_{\rm B}$/Ce.  
These values are close to those reported in CeRh$_{1-x}$Ir$_x$In$_5$~\cite{chr05}\@. 
%%%
It should be emphasized here, however, that there is {\it a crucial difference that the incommensurate AFM peak is not observed for $x \ge 0.4$}, within the experimental accuracy.
%However {\it the crucial difference is that the incommensurate AFM peak is not observed for $x$=0.4, 0.5, and 0.7}, within our experimental resolution; the integrated intensity of incommensurate AFM peak is, if any, less than 1/100 of that of commensurate AFM peak. 
%Since the magnetic neutron diffraction cross section is proportional to the square of the magnetic moment, the volume fraction of the incommensurate AFM phase is estimated to be less than 1/400 of that of the commensurate AFM phase.  

% Fig 2
\begin{figure}
  \includegraphics[scale=0.35]{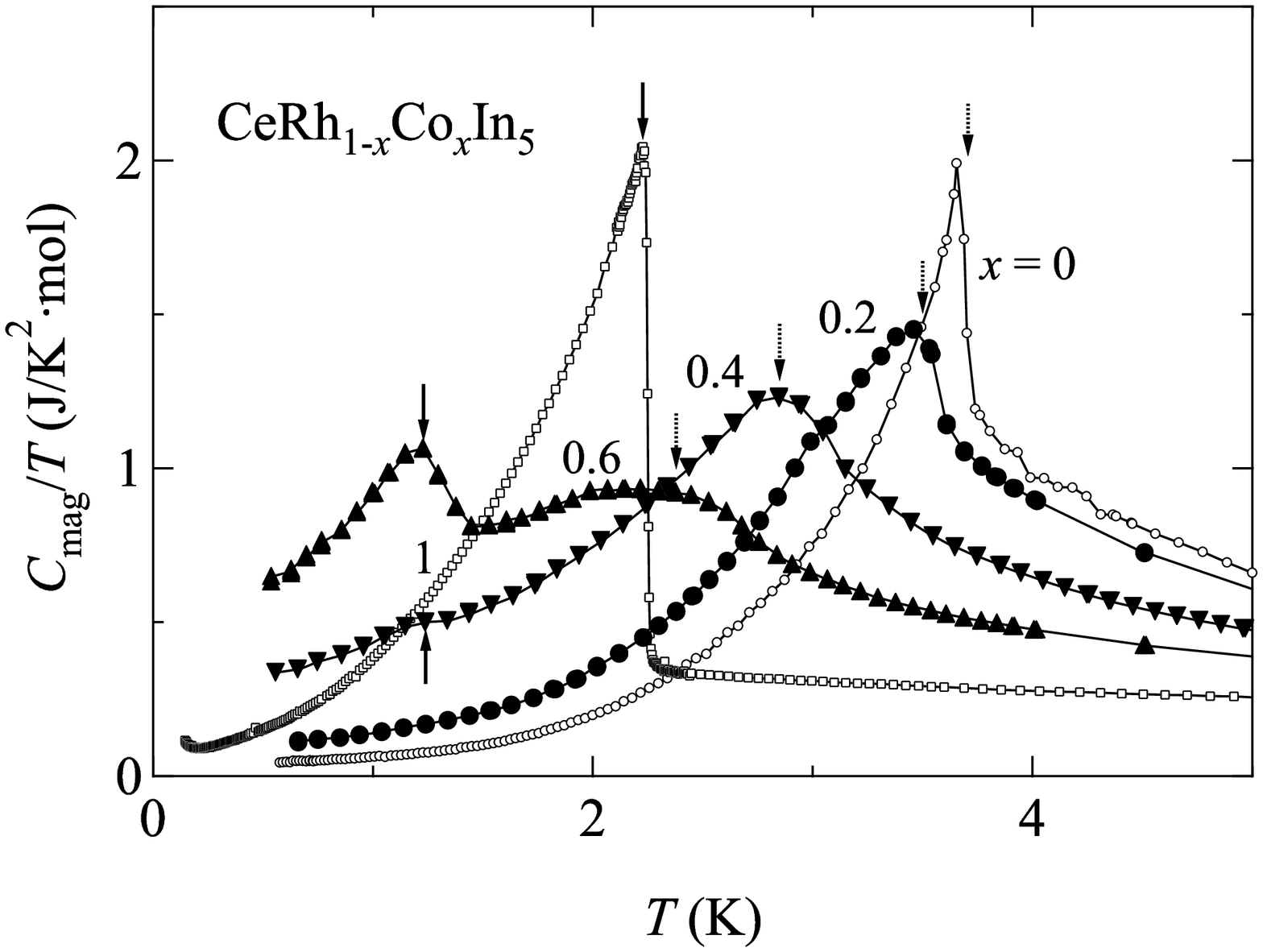}
  \caption{\label{specific_heat}
    Specific heat divided by temperature, $C_{\rm mag}/T$, of 
    CeRh$_{1-x}$Co$_x$In$_5$ as  a function of temperature. 
    The solid arrows indicate the SC transition, and the dotted ones
    the AFM transition.}
\end{figure}

%%% 2007-04-07 added
A further support of the absence of the incommensurate AFM order at $x \ge 0.4$ is provided by specific heat measurements which probe the bulk thermodynamic properties. 
Figure~\ref{specific_heat} shows the temperature dependence of the magnetic specific heat divided by temperature, $C_{\rm mag}/T$, 
for $x=0$, 0.2, 0.4, 0.6 and 1\@. 
Here $C_{\rm mag}/T$ is obtained by subtracting nonmagnetic contributions estimated by $C/T$ of LaRhIn$_5$\@.
In $x = 0.4$, two anomalies of the specific heat associated with the commensurate AFM and SC transitions are observed at $T=2.9$~K and 1.2~K, respectively, and no further anomaly is not observed. 
These results are consistent with the present neutron diffraction measurements.
%The jump on $C_{\rm mag}/T$ at $T_{\rm c}$ in $x = 0.4$ is much smaller than those in the  previous reports~\cite{chr05,yok06} and that observed in pure CeCoIn$_5$\@.
%However, this behavior is similar to those observed in CeRhIn$_5$ under pressures. 
%where the superconductivity coexists with the incommensurate AFM order~\cite{par06,kne06}.

\begin{figure}
  \includegraphics[scale=0.62]{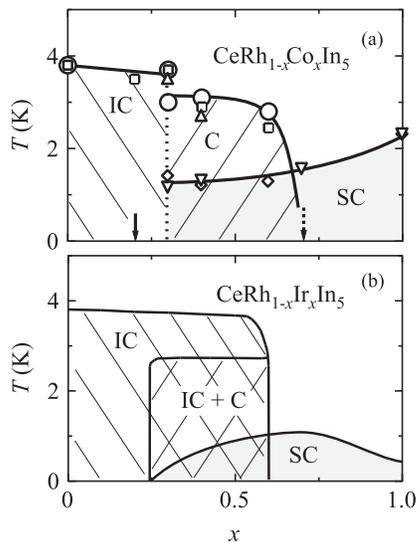}
  \caption{\label{diagram}
    (a)~$x$-$T$ phase diagram for CeRh$_{1-x}$Co$_x$In$_5$ from 
    the neutron diffraction ($\bigcirc$), specific heat ($\square, 
    \diamond$) and resistivity measurements ($\bigtriangledown,
    \triangle$)\@.
    At $x$=0.2 (solid arrow) and $x$=0.7 (dotted arrow),
    the superconductivity and the AFM order were not observed down
    to $T$=0.7~K, respectively.
    (b)~$x$-$T$ phase diagram for CeRh$_{1-x}$Ir$_x$In$_5$ 
    reported in refs.~\onlinecite{chr05} and \onlinecite{kaw06}
    is illustrated schematically.
  }
\end{figure}

The $x$-$T$ phase diagram for CeRh$_{1-x}$Co$_x$In$_5$ determined by the present neutron diffraction, specific heat and resistivity measurements is depicted in Fig.~\ref{diagram}(a)\@. 
%For $x=0.3$, 0.4, 0.6, and 0.7, the SC transition temperatures are determined by the specific heat and resistivity measurements. 
The incommensurate AFM order, which is observed in the pure CeRhIn$_5$ system, appears below $x=0.3$ and is absent at $x \ge 0.4$\@. The SC state is not observed down to 0.7~K at $x = 0.2$, while it suddenly appears at $x\sim 0.3$\@.
The commensurate AFM order simultaneously appears here, and stays on the intermediate $x$ region ($0.3 \le x \le 0.6$), together with the superconductivity. 
%At $x = 0.2$, no SC transition is observed down to 0.7~K\@.   At $x\sim 0.3$, the superconductivity suddenly appears.  Concomitantly, the commensurate AFM order appears.  The coexistence of superconductivity and commensurate AFM order is observed in the range $0.3 \alt x \alt 0.6$.  The commensurate AFM order appears to disappear at $x\sim$ 0.6\@. 

We now compare our phase diagram (Fig.~\ref{diagram}(a)) with that of CeRh$_{1-x}$Co$_x$In$_5$ reported previously.
The composition ($x$) dependence of $T_c$ is consistent with the
results reported in ref.~\onlinecite{jef05}, but the $x$ dependence of the magnetic transition temperature shows a clear difference. Namely, the present results show step-like behavior in contrast to a smooth curve from $x=0$ to the QCP ($x \sim 0.75$) in ref.~\onlinecite{jef05}\@.
At the present stage, we do not know the origin of this difference. 
Judging from the disappearance of both the superconductivity and the commensurate AFM order at $x=0.3$ and the discontinuity of $T_{\rm N}$, however, 
we conclude that the phase boundary between the incommensurate and
commensurate AFM phases is of first order, and then
the coexistence of the commensurate and incommensurate AFM orders observed at $x=0.3$ is attributed to small inhomogeneity in the composition at the first order phase boundary. 
%%% added
A similar coexistence reported in ref.~\onlinecite{yok06} for $x=0.4$ might be due to the similar inhomogeneity, as was occurred in our $x=0.3$ sample.

For further comparison, we schematically illustrate the phase diagram of the related material CeRh$_{1-x}$Ir$_x$In$_5$ reported in refs.~\onlinecite{chr05} and \onlinecite{kaw06}, in Fig.~\ref{diagram}(b)\@.
The two phase diagrams shown in Fig.~\ref{diagram} bear some resemblance; First, simultaneous appearance of the superconductivity and commensurate AFM order is observed at low $x$ regime. Second, the superconductivity coexists with the commensurate AFM order in the intermediate $x$ regime. 
%Thirdly, disappearance of the commensurate AFM order and concomitant increase of the SC transition temperature is observed at Co-rich side.  
However, a significant difference also exists there. Namely, while the incommensurate AFM order coexists with the superconductivity in CeRh$_{1-x}$Ir$_x$In$_5$, there is no intrinsic coexistence of the incommensurate AFM order with the commensurate AFM order and the superconductivity in CeRh$_{1-x}$Co$_x$In$_5$, implying that the superconductivity is strongly suppressed by the incommensurate AFM order in the latter system.
%it lead to a strong suppression of the superconductivity in CeRh$_{1-x}$Co$_x$In$_5$. 
A possible origin for this will be discussed later.

It may also be meaningful to compare the present results to some other experimental results on CeRhIn$_5$ under pressure. Recently, specific heat measurements under hydrostatic pressure revealed that the incommensurate AFM order suddenly disappears above a critical pressure $ p_c^* \sim 2$~GPa where a bulk SC phase sets in~\cite{kne06}\@.
%We also comment on the neutron results of CeRhIn$_5$ under pressure up to 1.63~GPa, in which superconductivity coexists with the incommensurate AFM order.  In ref. [\onlinecite{llo04}], the SC transition is determined by the resistivity measurements.  However the pressure range at which the bulk superconductivity is observed by the heat capacity measurements is higher than that determined by the resistivity measurements.  We therefore point out that the pressure range in the neutron measurements in ref. is not high enough to conclude the coexistence of bulk superconductivity and incommensurate AFM order. 
On the other hand, very recent NQR experiments have revealed a magnetic transition from incommensurate to commensurate at 1.67~GPa\@. The superconductivity coexists with the commensurate AFM order, and $T_c$ steeply increases above it~\cite{yas07}\@. 
%%% added
Therefore, the absence of the coexisting phase of the incommensurate AFM order and the superconductivity seems to be a common feature in the CeRh$_{1-x}$Co$_x$In$_5$ system and CeRhIn$_5$ under pressure.

The present result that, in CeRh$_{1-x}$Co$_x$In$_5$, the superconductivity competes with the incommensurate AFM order but
coexists with the commensurate one can provide an important insights for the mechanism of the unconventional superconductivity in this system.  
Namely, the area of the Fermi surface which disappears by the gap
formation due to the incommensurate AFM order plays an active
role for the superconductivity. 
%because the order leads to a strong
%suppression of the superconductivity.   On the other hand, 
However, the area which disappears at the commensurate AFM order may not be
important for the superconductivity, because the superconductivity
coexists with the commensurate AFM order.
%the area of the Fermi surface which disappears by the gap formation caused by the incommensurate AFM order appears to play an active role for the superconductivity, because the order leads to a strong suppression of the superconductivity.   On the other hand, the area which disappears at the commensurate AFM order may not be important for the superconductivity, because superconductivity coexists with the commensurate order.

\begin{figure}
  \includegraphics[scale=0.30]{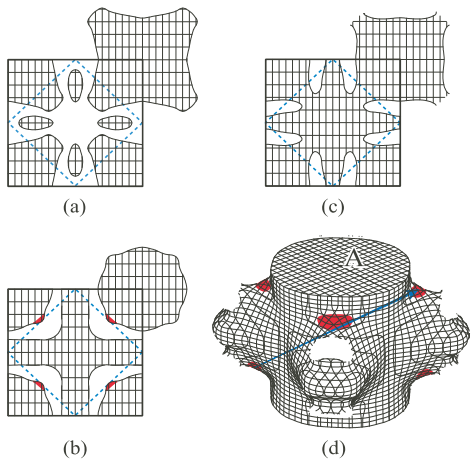}
  \caption{\label{fermi_surface}
    (a)-(c)~The cross section of the 14th Fermi surface perpendicular
    to $k_z$, at (a)~$k_z=0.149$, (b)~0.351, and (c)~$\frac{1}{4}$\@. 
    The blue dotted lines represent the boundary of the AFM Brillouin zone. 
    (d)~The 3D figure of the cylindrical part of 14th Fermi surface. 
    The red hatched regions represent the area which connected by 
    {\boldmath $q$}$_I=(\frac{1}{2},\frac{1}{2},0.298)$. 
    The arrow indicates the (0~0~1)$-${\boldmath $q$}$_I$ vector. 
    The red area is suggested to play an active role for the SC
    pairing formation. }
\end{figure}

Then it is tempting to discuss which area on the Fermi surface is
connected by the {\boldmath $q$}$_I$- and {\boldmath $q$}$_c$- wave
numbers.   According to the de Haas-van Alphen experiments, the 14th
band has the heaviest mass~\cite{set01}\@.  
%%%
We therefore assume that the 14th band is the main band for the
superconductivity. 
%%%
%%% in CeCoIn$_5$?
%%% Check ref. [R. Settai et al.] !!   
%%%
It is necessary to search all nesting positions connected by {\boldmath $q$}$_I$ and {\boldmath $q$}$_c$ in the 3D Fermi surface, but we discuss here the area symmetric about the $\Gamma$-point for simplicity.  Figures~\ref{fermi_surface}(a), (b), and (c) illustrate the cross section of the 14th band perpendicular to $k_z$ at $k_z$=0.149, 0.351 and $\frac14$, respectively.  The distances between the pairs of these sections inverted with respect to the $\Gamma$-point equal to the $z$-components of {\boldmath $q$}$_I$, (0~0~1)$-${\boldmath $q$}$_I$, and {\boldmath $q$}$_c$, respectively. 
%Then the blue lines shown in Figs.4 (a) and (b) are the connected by {\boldmath $q$}$_I$, while those shown in Fig.~4(c) are connected by {\boldmath $q$}$_c$.  
The blue dotted lines represent the boundaries of the Brillouin zone when the AFM orders set in.  Thus the positions of the Fermi surface which intersect the blue lines are supposed to be strongly influenced by the AFM orders.  At a first glance, the area at the corrugation on the cylindrical Fermi surface painted in red in Fig.~\ref{fermi_surface}(b) appears to be strongly nested through the (0~0~1)$-${\boldmath $q$}$_I$ vector. 
%%%
Taking into account the fact that the disappearance of the
superconductivity is concomitant with the appearance of the incommensurate AFM order,
it is natural to interpret that this area of the Fermi surface plays
active roles for the occurrence of both the superconductivity and
the incommensurate AFM order, 
and so that the gap formation accompanied with the incommensurate AFM order strongly suppresses the superconductivity. 
%Therefore it is natural to assume that this area plays an active role for the occurrence of the superconductivity.  
On the other hand, such a large nesting area is absent at $k_z =0.25$, as shown in Fig.~\ref{fermi_surface}(c)\@.  This seems to be relevant to the fact that the superconductivity is insensitive to the commensurate AFM order.   Figure~\ref{fermi_surface}(d) illustrates the cylindrical part of the 14th band.   The red area connected by 
(0~0~1)$-${\boldmath $q$}$_I$ (blue arrow) plays an active role for the superconductivity.

We finally discuss the difference between CeCoIn$_5$ and CeIrIn$_5$ inferred from the present study.   The most remarkable difference is that the incommensurate AFM order strongly suppresses the superconductivity in CeRh$_{1-x}$Co$_x$In$_5$,  while they coexist in  CeRh$_{1-x}$Ir$_x$In$_5$\@.  This implies that the active area on the Fermi surface for the superconductivity is different in these two systems.  Interestingly, a possible difference of the SC gap symmetry in these two systems has been suggested very recently by the thermal conductivity measurements: the line nodes are located perpendicular to the $ab$-plane in CeCoIn$_5$, while parallel to the $ab$-plane in CeIrIn$_5$~\cite{sha06}\@.

In summary, the neutron diffraction measurements reveal that, in CeRh$_{1-x}$Co$_x$In$_5$, the superconductivity competes with the incommensurate AFM order, while it coexists with the commensurate one.  This is in sharp contrast to  CeRh$_{1-x}$Ir$_x$In$_5$ system, in which both the commensurate and incommensurate magnetic orders coexist with superconductivity.   Based on these results, it is suggested that particular positions on the Fermi surface nested by (0~0~1)$-${\boldmath $q$}$_I$ may play an active role in forming the SC state in CeCoIn$_5$\@.  The present results further imply that the incommensurate spin fluctuation originating from the nesting characterized by (0~0~1)$-${\boldmath $q$}$_I$ plays an important role for the pairing interaction. To confirm this, the neutron quasi- and in-elastic scattering experiments through the SC transition is strongly desired.

We thank H.~Amitsuka, N. Furukawa, H.~Ikeda, H.~Kontani, S.~Onari, M.~Sigrist and M.~Yashima for valuable discussion, and M.~Azuma, Y.~Onuki, R.~Settai and Y.~Shimakawa for assistance in some of the measurements.  This work was partly supported by a Grant-in-Aid for Scientific Research from the Ministry of Education, Culture, Sports, Science and Technology.  H. S. was supported by the Research Fellowships of the Japan Society for the Promotion of Science for Young Scientists.

%\begin{acknowledgments}
%Acknowledgments
%\end{acknowledgments}

%They turn out to be Eqs.~(\ref{appa}), (\ref{appb}), and (\ref{appc}).
%\newpage %Just because of unusual number of tables stacked at end

\end{document}